# CLOUD COMPUTING SECURITY IN BUSINESS INFORMATION SYSTEMS

Sasko Ristov, Marjan Gusev and Magdalena Kostoska

Faculty of Information Sciences and Computer Engineering, Ss. Cyril and Methodius University, Skopje, Macedonia
{sashko.ristov, marjan.gushev, magdalena.kostoska}@finki.ukim.mk

## ABSTRACT

*Cloud computing providers' and customers' services are not only exposed to existing security risks, but, due to multi-tenancy, outsourcing the application and data, and virtualization, they are exposed to the emergent, as well. Therefore, both the cloud providers and customers must establish information security system and trustworthiness each other, as well as end users. In this paper we analyze main international and industrial standards targeting information security and their conformity with cloud computing security challenges. We evaluate that almost all main cloud service providers (CSPs) are ISO 27001:2005 certified, at minimum. As a result, we propose an extension to the ISO 27001:2005 standard with new control objective about virtualization, to retain generic, regardless of company's type, size and nature, that is, to be applicable for cloud systems, as well, where virtualization is its baseline. We also define a quantitative metric and evaluate the importance factor of ISO 27001:2005 control objectives if customer services are hosted on-premise or in cloud. The conclusion is that obtaining the ISO 27001:2005 certificate (or if already obtained) will further improve CSP and CC information security systems, and introduce mutual trust in cloud services but will not cover all relevant issues. In this paper we also continue our efforts in business continuity detriments cloud computing produces, and propose some solutions that mitigate the risks.*

## 1. INTRODUCTION

Cloud computing is not only the technology or concept but is already a reality. The number of cloud service providers and the number of various cloud services are increasing. Not only world's leaders in IT, but infants as well, offer cloud services (Iaas, PaaS, SaaS), such as Amazon's AWS and EC2, Google's App Engine, Salesforce's Sales and Service Cloud, Microsoft's Azure and Live, IBM SmartCloud etc, for various type of consumers: cloud service providers (CSPs), cloud consumers (CCs), and end users (EUs) (Table 1). Growth of various cloud service offer increases the number of cloud service consumer.

Cloud computing offers a lot of benefits, such as cost saving, on-demand services, scalability, redundancy and elasticity, and security benefits, as well. Furthermore, the cloud computing concept offers several business continuity benefits: eliminating downtime, better network and information security management, disaster recovery with both backup management and geographical redundancy [16]. It also avoids or eliminates disruption of operations, increases service availability and mitigates DoS attack possibility.

Despite the benefits, cloud computing produces many open security issues to be assessed. Migrating companies' services into cloud means that their data and applications are moving outside of the company security perimeter. This outsourcing opens new security issues and amplifies existing, thus increasing the company's security overall risk. Multi-tenancy, supported by virtualization, is another important security flaw producing new threats and vulnerabilities from inside, the co-tenants. The current isolation facility within clouds i.e. virtualization is weak and can be easily attacked [5]. The problem is even worse in the case of tenants are hosted on the same physical hardware. Thus, CSPs and CCs must ensure the customers that their data and applications are "really" secured and the risks are mitigated to the customer's acceptable level. A good discussion about securely exploitation of the benefits that cloud computing provides is given in [17].

Table 1. Cloud platform offers for different cloud service layers

| Service Layer | Amazon | Salesforce | Microsoft | Google | IBM |
|---|---|---|---|---|---|
| IaaS | EC2, S3, Simple Queue Service, SimpleDB | | | | SmartCloud |
| PaaS | | Force.com, Heroku, Database.com | Azure (Windows, SQL, .NET) | Google App Engine | CloudBurst Appliance |
| SaaS | | Sales cloud, Service Cloud | Live, Hotmail, Office Web App | Gmail, Google Docs | Lotus Live, Blueworks Live |

There are various existing international standards and "best practice" documents covering the security area. Obtaining ISO/IEC 27001 [3] or ISO/IEC 27002 [18], NIST-FISMA [21], or SSAE 16 (previously SAS 70) certification [22], or achieving PCI DSS [23] or HIPAA [24] compliance, can help CSPs to improve CCs' and EUs' trustworthiness in their cloud platforms' security. However, these standards are still far from covering the full complexity of the cloud computing model [4]. For example, ISO/IEC 27001 requirements are generic and are intended to be applicable to all organizations, regardless of type, size and nature. Other documents, [15,25,26,27] contain assessment issues directly related to cloud security, but do not address their implementation. Thus, it is necessary to assess the CSPs overall security, their obligations and responsibilities in order to ensure minimum the same, or even better, level of security than legacy CC's on-premise traditional solutions.

A brief overview about existing international security standards with cloud computing security with our evaluation of top CSP's security certification and assessments is presented in Section 2 (a more detailed in Appendix I). In Section 3 we propose a new methodology to evaluate ISO 27001:2005 control objectives. For this purpose we realize quantitative and qualitative comparative analysis of the ISO 27001:2005 requirement's importance if the company's services are hosted on-premises or at CSP. Next, we analyze the ISO 27001:2005 requirements' conformity to cloud computing security challenges due to cloud computing multi-tenancy, virtualization, and outsourcing the CCs' data and applications. Finally, despite the benefits, we evaluate business continuity detriments that cloud platform produces, and we introduced proposals which minimize the impact to business continuity and the probability of incident scenario for each detriment.

## 2. SECURITY STANDARDS AND ASSESSMENTS OVERVIEW

Security and privacy assessments are considered as best practice for evaluating a system or application for potential risks and exposures [2]. Traditional security assessments for on-premise infrastructure and applications and compliance audits are well defined and supported by multiple standards. But, additional challenges arise when audit tools are used to audit cloud environments [14].

In this section we briefly overview current existing international and industry standards, guidance, and best practices towards security.

The ISO/IEC 27000 series of standards, especially ISO/IEC 27001:2005 [3] for information security management system (ISMS), ISO/IEC 27002:2005 [18] for developing effective security management practices, and ISO/IEC 27005:2011 [1] for information security risk management (ISRM) have been developed as general purpose standards, and thus should be relevant for CSPs.

Another security control based guidance is NIST's special publication 800-53 R3 [28], as well as NIST's special publication 800-39 [29] for risk management at the organizational level.

COBIT [31] and SAS 70 (Audit) Type II (now replaced by two new standards) are audit and assessment standards and guidance for internal / external audits prior certification.

Despite general purpose standards, there are other security standards that cover specific areas. HIPAA [24] addresses the security and privacy of health data. PCI DSS V2.0 [23] is developed to encourage and enhance cardholder data security and facilitate the broad adoption of consistent data security measures globally.

We continue with overview of the existing CSPs' security certification and accreditations, as well as their security features. Table 2 presents the evaluation of the security standards certification that existing CSPs have, as well as their security features. As shown, all CSPs have one or many security certificates or compliances for their infrastructure. In addition, many CSPs not only they implement security features in their cloud systems, but they also offer security features to their customers to assess whether their services hosted in the CSP cloud are compliant to particular security standard.

Table 2. Existing CSPs' Security Certification and Accreditation, as well as Security Features

|  | **Amazon** | **Salesforce** | **Microsoft** | **Google** | **IBM** |
|---|---|---|---|---|---|
| **Security Compliance** | PCI DSS Level 1, ISO 27001, SAS 70 Type II, HIPAA | ISO 27001, SysTrust, SAS 70 Type II | PCI DSS, HIPAA, SOX,ISO 27001, SAS 70 TYPE 1 and II | SAS 70 Type II, FISMA | ISO 27001 |
| **Security Features** | AWS IAM, AWS MFA, Key Rotation | System status, Management Commitment for privacy | Access control, segmenting customer data | 2-step verification | Rational AppScan OnDemand, Security Services for compliance |

# 3. A New Methodology to Evaluate ISO 27001:2005 Control Objectives in All Cloud Computing Service Layers

We propose a model to measure the ISO 27001:2005 control objectives' importance factor for both on-premise and cloud solution, due to ISO 27001:2005's generality and because of almost all main CSPs' are ISO 27001:2005 Certified (Table 2). We assess and assign a quantitative metric for each control objective's importance factor, with details in IaaS, PaaS and SaaS cloud service layers. With the qualitative and quantitative analysis we compare the applicability and importance of ISO 27001:2005 control objectives as a general purpose standard, and the fact that the cloud techniques subsume the on-premise ones.

## 3.1 Metric Definition

In this paper we introduce a new methodology to determine appreciation / depreciation of control objectives' importance factor for defined control objectives during the processes of establishing the ISMS for ISO certification candidates, or during the processes of reviewing ISMS and its improvement, if the customer's resources are hosted on-premise or in the cloud. The indicators are defined by 6 possible values for the importance factor of each control objective. Table 3 shows the explanation of each importance factor value. The value "-" means that a particular control objective has no effect whether the services are hosted on-premise or in the cloud and it is irrelevant for a particular operational or management control objectives. Values from 1 to 5 mean that a particular control objective has different importance factor value if the services are hosted on-premise or cloud according to the given explanation.

Table 3. Control objective importance metrics

| value | Importance factor |
|-------|-------------------|
| - | Irrelevant if service is hosted on-premise or cloud |
| 1 | Minimal importance (most part moved to SLA) |
| 2 | Partial importance |
| 3 | Important |
| 4 | High importance (almost always) |
| 5 | Highest importance (important for each company / IS) |

## 3.2 Control Objectives Importance Factor Evaluation

Comparing the differences among three service layers of cloud computing versus traditional on-premises computing can be carried through deducing which resources or services are executed by CC and CSP. We can see such comparison on Figure 1 [12]. The resources and services in responsibility of the CSP are shown in green boxes, while those in responsibility of the CC are shown in red. Figure 1 shows that SaaS solution may be used from anywhere and at any time, provided a client (web browser) and internet connection. These features make SaaS software most attractive for SMEs, and require no additional expensive and complex resources and hardware on customer's part.

Figure 1 clearly presents that the responsibilities for all parts of the IT services hosted on-premises are on the resource owner, the customer in our case. Going from IaaS, through Paas to SaaS service layer in the cloud computing, more and more responsibilities are transferred from the CC to CSP. Therefore, CCs should transfer the security responsibilities to CSP, as well, and thus, most part of ISO 27001:2005 control objectives shall depreciate their values going from On-premise, to Iaas, Paas or SaaS.

Using this comparison, and according to defined metrics in Table 3, we evaluate each control objective importance factor on-premise and in IaaS, PaaS and SaaS cloud service layers. At the beginning, using control classification in [28] for control objectives, that is, (1) Management, (2) Operational and (3) Technical class, we grouped the ISO 27001:2005 control objectives. For (1), management control objectives, we expect that their importance factor does not depend if the company services are hosted on-premise or in cloud. For example, the company must define security policy, no matter of information systems' size, type, hosting and number. For (2), Operational control objectives, we expect that their importance factor should be depreciated if hosted in cloud, due to cloud benefits, such as redundancy, scalability, geographic spread, etc.

For (3), Technical control objectives, we also expect that their importance factor should be depreciated if services are hosted in cloud, due to technical benefits offered by the cloud.

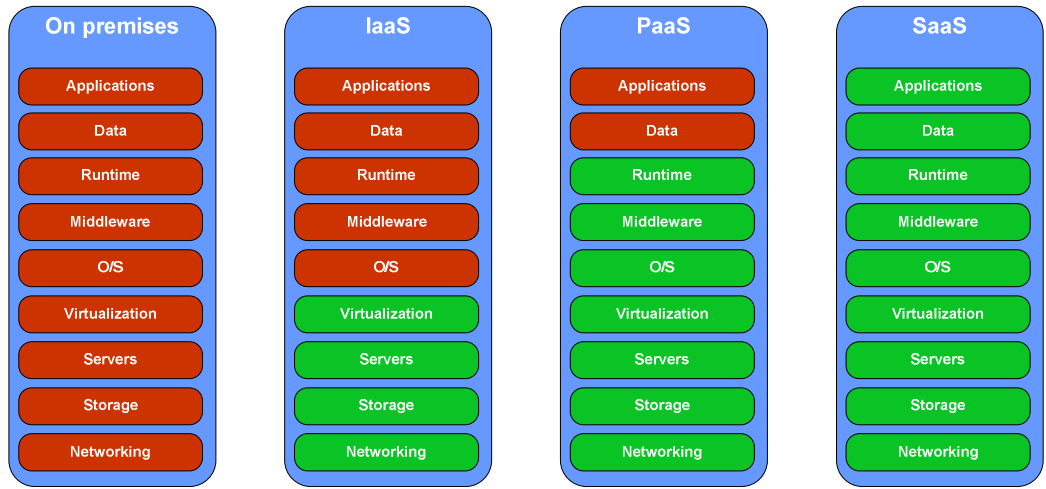

Figure 1. Comparison between On-premises computing, IaaS, PaaS and SaaS.

The summary results of the evaluation are presented in Table 4. The first two columns are the control objectives with their codes, as defined in ISO:IEC 27001:2005 [3]. The presented values in Table 4 are achieved on our evaluation of ISO/IEC 27001:2005's control objective importance factor, both for on-premise and the three cloud service layers, i.e. IaaS, PaaS and SaaS. Details of the evaluation are presented in Appendix II. As we can see, many control objectives depreciate their importance, but also, few other control objectives increase. We must emphasize that importance depreciation does not mean that a given control objective's meaning is decreased or even irrelevant or that particular control objective should be excluded, but the control objective obligations are somehow be transferred to the CSP, and should be integrated (partially or all controls of a given control objective) into SLA agreement signed between particular CSP and CC. During the processes of establishing the ISMS for ISO certification candidates, or during the processes of reviewing ISMS and its improvement, the prospective CC can use this evaluation to select / exclude the controls and control objectives to cover the identified requirements, and to put more effort and resources to control objectives with higher importance.

We must address that the evaluation is made for a Small and Medium Enterprises (SMEs), having their own information system, network and hardware equipment, since EU industry is mainly composed by SMEs [25].

### 3.3 Control Objectives Importance Analysis

After the evaluation, we proceed to qualitative and quantitative analysis on the results of the evaluation. In the quantitative analysis, we analyze the number of control objectives which importance increased / decreased for on-premise, as well as for each cloud computing service (IaaS, PaaS, SaaS) and their average. From the results shown in Figure 2, we conclude that for each cloud computing service, the number of control objectives with depreciated importance for each cloud service layer (IaaS, PaaS, SaaS) is four to ten times greater than the number of the control objectives with increased importance when moving into the cloud.

Table 4. Evaluation of importance factor for ISO 27001:2005 control objectives.

| # | Control Objective | On-premise | SaaS | IaaS | PaaS | Avg |
|---|---|---|---|---|---|---|
| 5.1 | Information security policy | - | - | - | - | - |
| 6.1 | Internal organization | - | - | - | - | - |
| 6.2 | External parties | 1 | 5 | 5 | 5 | 5 |
| 7.1 | Responsibility for assets | 5 | 4 | 4 | 3 | 4 |
| 7.2 | Information classification | 5 | 4 | 4 | 3 | 4 |
| 8.1 | Prior to employment | - | - | - | - | - |
| 8.2 | During employment | - | - | - | - | - |
| 8.3 | Termination or change of employment | - | - | - | - | - |
| 9.1 | Secure areas | 4 | 2 | 2 | 2 | 2 |
| 9.2 | Equipment security | 4 | 3 | 3 | 3 | 3 |
| 10.1 | Operational procedures and responsibilities | - | - | - | - | - |
| 10.2 | Third party service delivery management | 1 | 5 | 5 | 5 | 5 |
| 10.3 | System planning and acceptance | 4 | 3 | 2 | 1 | 2 |
| 10.4 | Protection against malicious and mobile code | 4 | 4 | 3 | 1 | 3 |
| 10.5 | Back-up | 5 | 4 | 3 | 1 | 3 |
| 10.6 | Network security management | 5 | 2 | 2 | 1 | 2 |
| 10.7 | Media handling | 5 | 4 | 4 | 3 | 4 |
| 10.8 | Exchange of information | - | - | - | - | - |
| 10.9 | Electronic commerce services | 5 | 4 | 3 | 1 | 3 |
| 10.10 | Monitoring | 5 | 4 | 3 | 1 | 3 |
| 11.1 | Business requirement for access control | - | - | - | - | - |
| 11.2 | User access management | 5 | 4 | 3 | 2 | 4 |
| 11.3 | User responsibilities | - | - | - | - | - |
| 11.4 | Network access control | 5 | 4 | 3 | 1 | 3 |
| 11.5 | Operating system access control | 5 | 5 | 5 | 1 | 5 |
| 11.6 | Application and information access control | 5 | 5 | 5 | 5 | 5 |
| 11.7 | Mobile computing and teleworking | 4 | 5 | 4 | 1 | 3 |
| 12.1 | Security requirements of information systems | - | - | - | - | - |
| 12.2 | Correct processing in applications | - | - | - | - | - |
| 12.3 | Cryptographic controls | 3 | 4 | 4 | 1 | 3 |
| 12.4 | Security of system files | 5 | 4 | 3 | 1 | 3 |
| 12.5 | Security in development and support processes | 4 | 4 | 4 | 1 | 4 |
| 12.6 | Technical Vulnerability Management | 4 | 3 | 2 | 1 | 3 |
| 13.1 | Reporting information security events and weaknesses | 4 | 3 | 2 | 1 | 2 |
| 13.2 | Management of information security incidents and improvements | 5 | 5 | 5 | 5 | 5 |
| 14.1 | Information security aspects of business continuity management | 5 | 5 | 5 | 5 | 5 |
| 15.1 | Compliance with legal requirements | 5 | 5 | 5 | 5 | 5 |
| 15.2 | Compliance with security policies and standards, and technical compliance | 4 | 3 | 2 | 1 | 2 |
| 15.3 | Information systems audit considerations | - | - | - | - | - |
| | TOTAL | 116 | 107 | 95 | 61 | 95 |

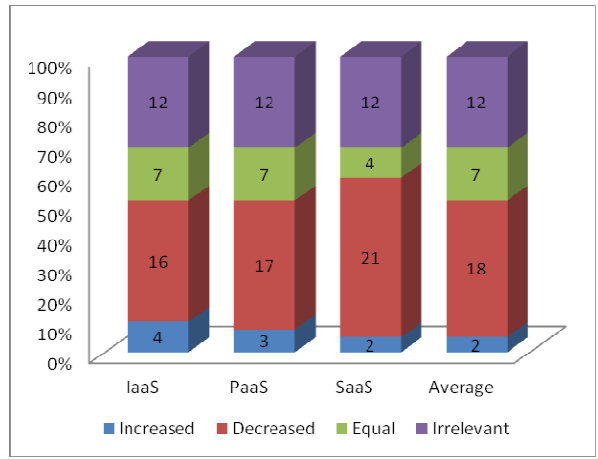

Figure 2. Comparison between On-premises computing versus cloud service layers - IaaS, PaaS and SaaS.

In the qualitative analysis, we analyze the sum of control objectives' importance factor value for each cloud service layer and the average, compared to the sum of the importance factor values of the same control objectives when hosted the services on-premise.

From the results, shown in Figure 3, we conclude that for each cloud computing service layer, total sum of control objective importance factor values into the cloud depreciates compared to on-premise, that is, before moving into the cloud. The percentages of the importance value depreciation are 7.76%, 18.10%, 47.41%, 18.10% for SaaS, IaaS, PaaS and Average, respectively.

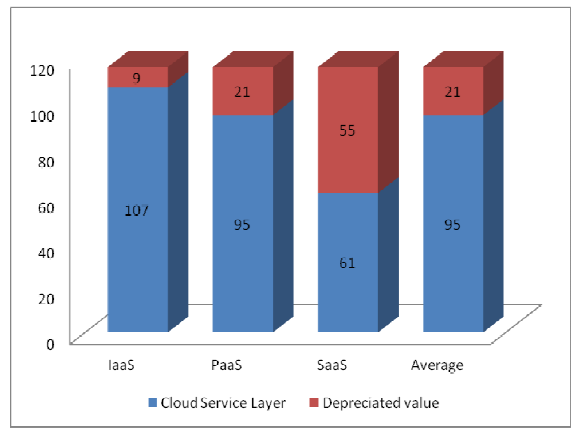

Figure 3. Qualitative analysis on ISO 27001:2005 control objectives when moving into each cloud service layer.

We can conclude that both quantitative and qualitative analysis result in the control objectives importance depreciation, in all three cloud service layers, as well as in their average. The depreciation percentages show some paradox, that is, the distribution of control objective importance is not equal to the responsibilities of CSP and CC, especially for SaaS, where it is expected to downgrade the importance near to zero. This is due to generality of ISO 27001:2005 and the fact that CCs will still have information and assets on-premise, employees, legal issues, etc.

# 4. INFORMATION SECURITY CONTROLS (IN)COMPLIANCE FOR CLOUD COMPUTING

In this Chapter we analyze the ISO 27001:2005 requirements' conformity to cloud computing security challenges, particularly on new one, such as customer isolation, insider attacks, and security integration [8], due to cloud computing multi-tenancy, virtualization, and outsourcing the CCs' data and applications. We evaluated that almost all main CSPs obtained ISO 27001:2005 certification (and others, as well). But, due to security challenges found, we analyze if CSP' ISO 27001:2005 Certificate will be enough to assure its CCs' that are secured in the rented infrastructure, platform or software.

## 4.1 Security Challenges due to Virtualization

Traditional on-premise data-centers security solutions do not comply with virtualized environment, because of the complex and ever-dynamic nature of cloud computing [7]. But, the virtualization by itself does not affect the security if it is used on-premise in a physical, logical and environmental isolated secured environment. IDS and IPS systems can secure the internal virtual and physical machines from the exterior environment, if they are into one autonomous system, that is, under same administrative governance.

In cloud computing, especially in IaaS and PaaS, the resources are shared and rented to the different customers. Even more, the same physical machine can be shared to many different customers. The current virtualization is weak and can be easily attacked [5]. The security solutions for some flaws are found, but new security threats and vulnerabilities arise day by day. Thus, CSPs' security perimeter is broken from inside, making their IDS and IPS helpless. Therefore, CSPs must introduce effective isolation among the CCs, although allowing physical resource sharing.

We found several security solutions to the virtualization challenges. Hao F. et al [10] propose SEC2 solution which enables users to customize their security policy settings the same way they control their on-premise network. Ibrahim A. et al [6] propose Virtualization-Aware Security Solution CloudSec, which monitors volatile memory to detect and prevent for the kernel data rootkits.

Analyzing ISO 27001:2005 requirements and their controls, we concluded that control that covers the virtualization does not exist. Indeed, in clause 11, that is, access control, many standard controls, even whole control objective, assumes that operating systems are on separate real machines. But, in the reality, issues such as trusting the VM image, hardening hosts, and securing inter-host communication are critical areas in IaaS [11]. Therefore, we propose to include new control objective for virtualization management, with two controls: *virtualization* and *virtual machines control*. For the former, we propose: *Information involved in virtual machines shall be appropriately protected*. For the latter, we propose: *Virtual machines shall be adequately managed and controlled, in order to be protected from internal and external threats, and to maintain information security in transit*. In addition to this, NIST defines control SC-30 in [28], *Virtualization Techniques*, which is not mapped to any control of ISO 27001:2005 [28]. But, NIST's control is far from enough to cover all security flaws due to virtualization, especially when it is gathered together with multi-tenancy.

## 4.2 Security, Data Protection and Privacy as-a-Service

Business does not fully accept cloud infrastructure, platform and software due to security, data protection and privacy, as well as trust issues. Combining the advantages of secured cloud storage and software watermarking through data coloring and trust negotiation, the authors in [9] propose reputation system to protect data-center access at a coarse-grained level and secure data access at a fine-grained file level.

Such systems and solutions supersede and subsume the traditional security systems, and thus CSPs should implement them. Therefore, offering *Security-as-a-Service* (*SECaaS*) and *Data protection and privacy-as-a-Service* will speed up cloud market growth, both for the providers' offers and clients, as well as cloud trustworthiness. CSA offers 10 candidate domains for SECaaS [14].

Data privacy is treated in two controls in ISO 27001:2005 requirements. The control 6.2.3 requires the client data privacy (CCs) and the control 15.1.4 requires from the CSP to ensure data privacy. These two controls obligate both the CCs and the CSPs to manage the data privacy with higher importance.

As shown in Table 2, many CSPs are not only complained to some security standards, but they offer services to the CC to help them in their security standard compliance, as well. Thus, the risks that arise from multi-tenancy and virtualization will be mitigated, and mutual trustworthiness will be established among CSPs, CCs and EUs.

## 4.3 Performance challenges

All mentioned cloud computing security solutions and techniques, and many other as well, degrade cloud services' performance. Implementing identity and access management, web and email security, intrusion management, [14], as well as monitoring systems, data coloring, and other traditional security services, such as web service security produce data overhead and system latency which must be considered due to their negative impact to server performance, and thereby to the system availability.

## 5. BUSINESS CONTINUITY CHALLENGES IN THE CLOUD

As the CSP becomes an external party that CC relies, it has to develop and approve Business Continuity Plan (BCP), mapped to the international standards, such as [1]. Business continuity and Disaster recovery are one of the domains for CSA's SECaaS [14], and therefore, we analyze the security detriments cloud computing offers to SME.

Despite the benefits the cloud provides [16] for CC's business continuity, it also comes with many security detriments producing risks which impact the business continuity. Some of the benefits produce detriments, as well. The authors in [13] discuss cloud computing security risks and enlightened steps to mitigate them. In this paper we extend the analysis and overview some of the main risks that impacts to the business continuity, together with some solutions that mitigates the risks to acceptable level.

***Data Privacy***. CSPs must ensure the CCs into their operations and privacy assurance. Privacy-protection mechanisms must be embedded in all security solutions. This risk directly impacts the regulatory compliance risk and company business reputation.

***Proposal***: Auditing and logging tenant's activities can reduce the risk of incidents, as well as including obligations in the SLA agreements. ISO 27001:2005 do have controls for audit and logging, but CSP must include the two controls we proposed previously as well.

***Regulatory and Standards Compliance***. Cloud Service Provider must provide the evidence that meets the standards and regulatory the company needs.

*Proposal*: The CSP should permit the regular audits by the CCs. The CC should assess the risks and include them into risk acceptance plan, if acceptable. If not, the services with unacceptable risks should stay in-house. ISO 27001 covers these issues (Legal, third party, etc) well in several controls.

*Loss of Control*. Company must transfer some of the control of the assets, application, etc to the cloud service provider.

*Proposal*: CCs must assure that their CSP can meet SLA requirements, and if not, they must assess the risks and include them into BCP. Also, we suggest to CSPs regulatory to obligate CCs to concern about security in SLA agreement.

*Data Location*. In some cases, the applications and data might be stored in countries, which can judiciary concern, and lead to regulatory incompliance.

*Proposal*: Keeping these application and data in-house, or in the hybrid cloud, with the appropriate SLA can reduce this risk to acceptable level.

*Heterogeneity, Complexity, Interoperability*. Business continuity depends not only on the effectiveness and correctness of system components, but also on the interactions among them. Subsystem component heterogeneity leads to difficult interoperability. Number of possible interactions between components increases the system failure probability. Complexity typically relates inversely to security, with greater complexity giving rise to vulnerabilities [27].

*Proposal*: Defining security standards for adapters, wrappers, transducers, and data transformation, as well as performance analysis can offer stable system solution and mitigate the risks.

*Data Protection*. Although replication produces security benefits in Disaster Recovery and system availability, it produces a security detriment. Thus, along with virtualization, complicates the access control management and data privacy.

*Proposal*: Outsourcing only noncritical applications and its data to cloud, if applicable, shall provide the client company with even better data protection and management compared to traditional solutions.

*Multi-tenant environment*. Although cloud can offer better protection and defense for the same cost than traditional solutions, it has a detriment as well. Thus, due to shared and multi-tenant environment, especially in the public clouds, other cloud tenants are threat from within the cloud, which is not the case in the traditional in-house solution, even if virtualization techniques are used.

*Proposal*: CSP should develop a methodology to evaluate the tenants and categorize them into 5 categories in the manner of trustfulness, such as, very trusted, trusted, medium, untrustable, and very untrustable, especially for IaaS and PaaS, where clients have more impact to its own security, but also can be threats to the other tenants.

*Disaster Recovery - RPO and RTO*. Although cloud can offer better RPOs and RTOs [16], maybe CSP had not define these objectives, or did define, but worse than CCs expect.

*Proposal*: The CCs must be ensured that the CSP's RPOs and RTOs are defined in compliance with its own, as well as the CSP can satisfy such defined requirements.

## 6. CONCLUSION

Business managers know that risks exist in spite of all the benefits of each new technology or business model offers. There are a lot of regulatory violation, security, trust and privacy issues. Thus, each company that dives ahead using the benefits of cloud computing, should evaluate the risks found if moving its services into the cloud, compare to if retain to the traditional solutions.

In this paper, we overview main international and industry standards towards security, and analyze their conformity to cloud computing. There are many different cloud security threats, vulnerabilities and control definitions, best practices, in order to standardize cloud security, as well.

ISO 27000 series (27001:2005, 27002:2005, and 27005:2011) of standards are defined as generic and they cover not only the technical solutions to technically identified threats and vulnerabilities, but take into account the operational, organizational and management vulnerability, as well. Due to its generality, as well as many open cloud security challenges, ISO 27001:2005 is not fully conformal with cloud information security system. Therefore, we propose a new control objective in ISO 27001:2005 requirements, *virtualization management*, with two controls covering *virtualization* and *virtual machines control*.

In this paper we define a methodology to quantify the ISO 27001:2005 Requirements grouped in control objectives, for on-premise and cloud. We evaluate that moving into cloud, 12 of 39 control objectives are for management, and are not affected if the services are on-premise or in cloud. Importance factor doesn't change on average seven Control Objectives, depreciates on 18, and increases on only two of them. Thus, moving into cloud, CCs (SME) transfer the importance of the security to its CSP, and expect that their data and applications to be secured. Therefore, and due to emergent security challenges that cloud computing produces, CCs must re-evaluate their BCPs.

No paper so far has presented business continuity aspects in detail of cloud computing and it challenged us to address the cloud computing model security detriments that depreciate the CC business continuity: data privacy and protection, regulatory and standards compliance, loss of control, data location, heterogeneity, complexity, and interoperability, multi-tenant environment, and disaster recovery - RPO and RTO compliance and effectiveness. In this paper we introduced proposals which minimize the impact to business continuity and the probability of incident scenario for each detriment. These main risks can be assessed appropriately and mitigated to the acceptable level by applying recommendations in these proposals according to matrix for risk level as a function of the business impact and probability of incident scenario [1].

## 7. APPENDIX I

In this Section we present existing international security standards and guidelines in details, as well as the efforts within cloud computing security.

### 7.1 ISO/IEC 27000 Standard series

The ISO/IEC 27000 series of standards have been specifically reserved by ISO for information security matters.

*ISO/IEC 27001:2005* certification, for ISMS, can be considered as the best solution for securing provider's information assets, as well as to establish customer's trust in their services. Microsoft proves to customers that information security is central to its cloud operations [19]. The standard adopts the "Plan-Do-Check-Act" (PDCA) model, which is applied to structure all ISMS processes. Applying this model ensures that CSP ISMS is established, implemented, assessed, measured where applicable, and continually improved. The standard defines a set of 133 controls grouped into 39 control objectives, grouped into 11 clauses. These control objectives and controls shall be selected as part of the process of establishing ISMS as suitable to cover the identified requirements. They are not exhaustive and additional control objectives and controls may also be selected, or some can be excluded, but the prospective CSP must justify the exclusion.

*ISO/IEC 27002:2005*, complementary with ISO 27001:2005, is a practical guideline for developing organizational security standards and effective security management practices and to help build confidence in inter-organizational activities.

*ISO/IEC 27005:2011* provides guidelines for ISRM in an organization, supporting the requirements of an ISMS according [3]. ISRM process consists of context establishment, risk assessment, risk treatment, risk acceptance, risk communication and risk monitoring and review.

## 7.2 NIST's 800-53 R3 Security Controls

The NIST's special publication 800-53 R3, Security Controls for Federal Information Systems and Organizations, is another security control based guidance. It provides guidelines for selecting and specifying security controls for information systems (ISs) supporting the executive agencies of the federal government to meet the requirements of FIPS 200 [30]. The guidance defines total of 205 controls, grouped in 17 families of security controls for an information system and one family of program management controls to manage information security programs.

The standard focuses on managing risks aroused from information systems, and the risk management at the organizational level is incorporated in NIST's Special Publication 800-39.

## 7.3 Audit and Assessment Standards and Guidance

To obtain ISO 27001:2005 Certification, a company must perform internal and external audits prior certification. There are several guidance and certifications for this purpose.

*COBIT 4.1*. Control Objectives for Information and Related Technology (COBIT), latest version 4.1, developed by Information Systems Audit and Control Association (ISACA), provides a set of 34 high-level control objectives, one for each of the IT processes, grouped into four domains: Plan and Organize, Acquire and Implement, Deliver and Support, and Monitor and Evaluate. The structure covers all aspects of information and the technology that supports it. By addressing these 34 high-level control objectives, the business process owner can ensure that an adequate control system is provided for the IT environment. Next COBIT version 5 is in preparation.

*SAS 70 (Audit) Type II*. SAS 70, developed by American Institute of Certified Public Accountants (AICPA), does not specify a pre-determined set of control objectives or control activities that CSP must achieve, but it provides guidance to enable an independent auditor to issue an opinion on a CSP's description of controls through a Service Auditor's Report. SAS70 Type II certifies that CSP had an in-depth audit of its controls (including control objectives and control activities), which should relate to operational performance and security to safeguard CCs data. This helps the CSP to build trust with its CCs. CCs, on the other hand, with the Service Auditor Report from their CSP(s), obtain valuable information regarding the CSP(s) controls and the effectiveness of those controls. The standard SAS70 is now divided into parts and replaced by two new standards: (1) SSAE No. 16 for Service Auditors and (2) Clarified

Auditing Standard for User Organizations. But, we analyzed SAS 70 because most of the main CSPs are SAS 70 compliance.

There are other security standards that cover specific areas. HIPAA [24] addresses the security and privacy of health data and intends to improve the efficiency and effectiveness of the health care system by encouraging the widespread use of electronic data interchange. PCI DSS V2.0 [23] is developed to encourage and enhance cardholder data security and facilitate the broad adoption of consistent data security measures globally. At high level it has 12 requirements to protect cardholder data, which may be enhanced with additional controls and practices to further mitigate risks at acceptable level.

## 7.4 Cloud Security Efforts on Standardization

Although general security standards can help CSPs in implementing information security system, more efforts are done to cloud security standardization. CSA identified top threats to cloud computing in [20]. In order to mitigate the risks of threats, ENISA identified and assessed the risk level as a function of the business impact and likelihood of the incident scenario [25].

NIST discusses the threats, technology risks, and safeguards for public cloud environments and provides the insight needed to make informed information technology decisions on their treatment [27]. The main emphasis is putted on security and data privacy.

The Cloud Security Alliance's initial report, V2.1 [33] contains a different sort of taxonomy based on 15 different security domains and the processes that need to be followed in an overall cloud deployment. But, new candidate domains are proposed for new version 3 [14], which are of the greatest interest to experienced industry consumers and security professionals. For each candidate domain are addressed core functionalities, optional features, services, addressed threats, and the challenges to be focused on.

CSA puts a lot of efforts in its CSA GRC project, as well [32]. A list of 98 controls grouped into 11 groups is defined in [15]. More important is that each control is mapped into compliant control of other security standards or best practices.

## 7.5 Which standard is appropriate for CSP's Information Security System?

The best solution for CSP's information security system is to cover and meet both the ISO standard and NIST guidance controls. But, the answer to the question is not so simple. In NIST's 800-53 [28] is shown that a small number of controls are not covered in the other standard. Also, neither NIST's 800-53 security control subsumes ISO 27001:2005, neither opposite. There are many security controls with similar functional meaning, but with different functionality. Other security controls with similar topics are addressed in the same control objective (ISO) or family (NIST), but has different context, perspective, or scope. Another problem is that some controls from one standard are spread in several controls in the other standard.

The standards differ in their purpose and applicability, as well. While ISO 27001:2005 is general purpose and applies to all types of organizations, NIST's 800-53 is applicable for information systems supporting the executive agencies of the federal government.

The main concern here is: are the controls of both standards applicable to CSP and all cloud service layers IaaS, PaaS and SaaS? Do they cover all the traditional security challenges, as well as newly opened security issues in cloud? Are there any security challenges in cloud computing not covered with these controls?

ISO 27001:2005 is a general purpose standard and therefore, its control objectives are conformable to CSP. But the question remains: Are they enough for CSP? Our further research is going into two directions: first, we measure the CC efforts to be taken for each ISO 27001:2005 control objective if their services are hosted on-premise or in the cloud. And second, we analyze if there should be any other security control or even more, control objective, to be included in the ISO 27001:2005 controls.

# 8. APPENDIX II

Table 5 presents the details of importance factor evaluation for each ISO 27001:2005 control objective. The importance factor depends on fact whether the company services are hosted on-premise or in cloud.

Table 5. Details of ISO 27001:2005 Control objectives importance evaluation.

| # | Control Objective |
|---|---|
| 5.1 | **Information security policy**:<br>The organization must define security policy, no matter if the services are hosted on-premise or in cloud. Therefore this control objective is not evaluated. |
| 6.1 | **Internal organization:**<br>The information security must be managed within the organization for both solutions. Therefore this control objective is not evaluated. |
| 6.2 | **External parties:**<br>The organization's information and information processing facilities are accessed, processed, communicated to, or managed by external parties in each cloud service layer and we evaluate each cloud service layer with 5. For on-premise hosting the organization does not use external party services and we evaluate with 1. |
| 7.1 | **Responsibility for assets:**<br>The organization must manage appropriate protection of organizational assets on-premise regardless of company type (evaluate with 5). In Iaas and PaaS the organization transfers some of the assets (reduced to importance 4), and in SaaS most of the application data are completely transferred (reduced to importance 3, as data field is green in Figure 1). However, many other assets like mobile phones, paper accounting documents, usbs, etc the organization must manage. |
| 7.2 | **Information classification:**<br>Each organization shall perform an appropriate level of protection to information on-premise (importance is 5). Some of the procedures for information handling are transferred to CSP for IaaS and PaaS, and more for SaaS as data field is green for SaaS. |
| 8.1 | **Prior to employment:**<br>The organization must ensure that employees, contractors and third party users understand their responsibilities, and are suitable for the roles they are considered for, and to reduce the risk of theft, fraud or misuse of facilities for both solutions. Therefore this control objective is not evaluated. |
| 8.2 | **During employment:**<br>Similar to 8.1 this control objective is not evaluated. |
| 8.3 | **Termination or change of employment:**<br>Similar to 8.1 this control objective is not evaluated. |
| 9.1 | **Secure areas:**<br>The organization must prevent unauthorized physical access, damage and interference to the organization's premises and information. Some organizations are |
| | tenants in secured areas and therefore we evaluate on-premise with 4. The organization transfers most of this facility to CSP (Networking, Storage and Servers are green in Figure 1) and the importance factor is 2 for all cloud service layers. |
| 9.2 | **Equipment security:**<br>Similar to 9.1, but some equipment stays on premise and thus the equipment security is important in all cloud service layer. |
| 10.1 | **Operational procedures and responsibilities:**<br>The organization must secure the information processing facilities for both solutions. Therefore this control objective is not evaluated. |
| 10.2 | **Third party service delivery management:**<br>This control objective is very important for cloud solution as the CSP is external party and in many cases customer depends on third party service delivery (CSPs' external parties like CSPs' Internet service providers, power supply, etc). A customer must implement the control objective requirements into its BCP. Importance factor for on-premise is evaluated with 1 and for cloud with 5 since the external parties on-premise become third party in each cloud solution. |
| 10.3 | **System planning and acceptance:**<br>For this control objective we put 4 for on-premise due to standard generality, i.e. not all organizations possess information systems or update them. Going from IaaS to SaaS the organization shall transfer to CSP with SLA the risk of systems failures. In IaaS the hardware resources are transferred to CSP, in PaaS operating systems and runtime, as well, and in SaaS the applications. Therefore we decrease the importance factor starting from 1 for on-premise and going from IaaS to SaaS. |
| 10.4 | **Protection against malicious and mobile code:**<br>Same as 10.3 the importance factor is evaluated to 4 for on-premise due to standard generality. Protection level of the integrity of software and information for On-premise and IaaS solutions is the same. In PaaS solution CSP shall have some procedures to protect the customer (decreased importance by -1), and in SaaS CSP have the full responsibility of the operating systems and the applications (transferred in SLA, i.e. importance factor is 1). |
| 10.5 | **Back-up:**<br>Similar evaluation as 10.4, except for on-premise where each organization regardless of its nature, type and size must perform backup on some assets (Legal requirements, Archive, Accounting etc). |
| 10.6 | **Network security management:**<br>The organization in on-premise solution must ensure the protection of information in networks and supporting infrastructure (importance factor is 5). IaaS and PaaS are evaluated with 2 as the organization transfers more of the responsibility to CSP (there is network traffic, such as VPN or remote control from the organization to CSP). SaaS is evaluated with 1 solution the organization transfers the responsibility completely to CSP as defined in SLA. |
| 10.7 | **Media handling:**<br>Some media are unnecessary for cloud solution, such as backup media. However, not all media are transferred to CSP; for IaaS and PaaS the organization shall prevent smaller number of media than on-premise (importance factor is high 4), and even less for SaaS (importance factor is 3). |
| 10.8 | **Exchange of information:**<br>The security of information and software exchanged within an organization and with any external entity is not affected if the services are on-premise or hosted in the cloud.<br>Therefore this control objective is not evaluated. |

| 10.9 | **Electronic commerce services:**<br>The value of importance factor for on-premise solution is 5 if the organization uses e-commerce services. The importance in IaaS and PaaS decreases and for SaaS solution the CSP has completely responsibility, similar to 10.5. |
|---|---|
| 10.10 | **Monitoring:**<br>The organization transfers the responsibility for monitoring to CSP going from IaaS to SaaS, similar to 10.9 |
| 11.1 | **Business requirement for access control:**<br>This is high-level control objective and access control policy shall be established regardless the solution. Therefore this control objective is not evaluated. |
| 11.2 | **User access management:**<br>The organization transfers more of the responsibility to CSP (in SLA) going from IaaS to SaaS similar to 10.5. The difference is in SaaS since we evaluate it with 2 because the organization has to manage physical access. |
| 11.3 | **User responsibilities:**<br>Users have the same responsibilities no matter where the information is. Therefore this control objective is not evaluated. |
| 11.4 | **Network access control:**<br>The organization transfers the responsibility for monitoring to CSP going from IaaS to SaaS and therefore this control objective is evaluated similar to 10.5. |
| 11.5 | **Operating system access control:**<br>For on-premise, IaaS and PaaS the organization has the responsibility for operating system access control (importance factor is 5). We evaluate the importance of SaaS solution with 1 as the organization transfers the responsibility to CSP as depicted in Figure 1. |
| 11.6 | **Application and information access control:**<br>For both the solutions it is the responsibility to the organization for application and information access control. Therefore we evaluate both solutions with importance 5. |
| 11.7 | **Mobile computing and teleworking:**<br>Hosting the services in IaaS increases the importance of this control objective. We evaluate this control objective with maximum importance 5 for IaaS. For PaaS CSP transfers come of the responsibilities to CSP and we decrease the importance. For SaaS this control objective shall be included in SLA and therefore we evaluate with1. |
| 12.1 | **Security requirements of information systems:**<br>Security must be an integral part of information systems regardless of the solution. Therefore this control objective is not evaluated. |
| 12.2 | **Correct processing in applications:**<br>The organization must prevent errors, loss, unauthorized modification or misuse of information in applications regardless of the solution. Therefore this control objective is not evaluated. |
| 12.3 | **Cryptographic controls:**<br>We evaluate on-premise with 3 due to standard generality. We evaluate IaaS and PaaS with increased importance 4 as the organization must implement cryptography to protect the confidentiality, authenticity or integrity of information. For SaaS the CSP has the responsibility. |
| 12.4 | **Security of system files:**<br>Similar to 11.4 |
| 12.5 | **Security in development and support processes:**<br>Similar to 11.5 |

| 12.6 | **Technical Vulnerability Management:**<br>The organization transfers the responsibility to CSP going from IaaS to SaaS, similar to 10.3. |
|---|---|
| 13.1 | **Reporting information security events and weaknesses:**<br>Similar to 12.6 |
| 13.2 | **Management of information security incidents and improvements:**<br>For both the solutions it is the responsibility to the organization for management of information security incidents with highest importance factor 5. |
| 14.1 | **Information security aspects of business continuity management:**<br>There are many business continuity benefits and detriments that the organization must evaluate in its BCP. Therefore both the solutions are evaluated with maximum importance factor 5. |
| 15.1 | **Compliance with legal requirements:**<br>For both the solutions it is the maximum responsibility to the organization to comply with the legal requirements. Both the solutions are evaluated with maximum importance 5. |
| 15.2 | **Compliance with security policies and standards, and technical compliance:**<br>The organization transfers the responsibility to CSP going from IaaS to SaaS for compliance with security policies and standards, and technical compliance, similar to 13.1. |
| 15.3 | **Information systems audit considerations:**<br>This is a high level organizational control objective and therefore this control objective is not evaluated. |

**Authors**

Sashko Ristov is a PhD student in the area of security and performance of cloud computing. Hi is a teaching and research assistant at the Faculty of Computer Science and Informatics, Skopje, Macedonia. His areas of interest include High Performance Computing, Computer Networks, Security and Cloud Computing.

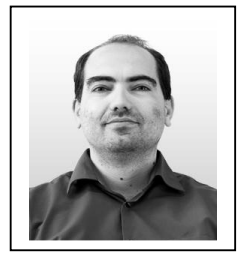

Marjan Gushev is a professor at the Faculty of Computer Science and Informatics, Skopje, Macedonia. He has published more than 40 papers in Parallel Computing related journals. His areas of interest include Computer architecture, Parallel Processing, Computer Networks, Internet, and Cloud Computing.

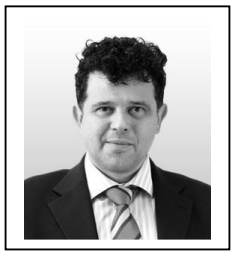

Magdalena Kostoska is a PhD student in interoperability in cloud computing. She is a teaching and research assistant at the Faculty of Computer Science and Informatics, Skopje, Macedonia. Her areas of interest include Cloud Computing, High Performance Computing, Data Structures, and Software Engineering.

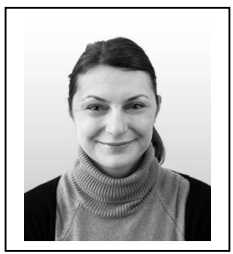